# Toward Exploratory Search in Biomedicine: Evaluating Document Clusters by MeSH as a Semantic Anchor

**Michael Segundo Ortiz[a], Kazuhiro Seki[b], Javed Mostafa[a,c]**

[a] *Carolina Health Informatics Program, University of North Carolina, Chapel Hill, NC, USA*
[b] *Department of Intelligence and Informatics, Konan University, Kobe, Hyogo, Japan*
[c] *School of Information and Library Science, University of North Carolina, Chapel Hill, NC, USA*

**Abstract**

*The current mode of biomedical literature search is severely limited in effectively finding information relevant to specialists. A potential approach to solving this problem is exploratory search, which allows users to interactively navigate through a vast document collection. As the first step toward exploratory search for specialists in biomedicine, this paper develops a methodology to evaluate quality of document clusters. For this purpose, we incorporate human expertise into data set creation and evaluation framework by leveraging MeSH terms as semantic anchors. In addition, we investigate the benefit of full-text data for improving cluster quality.*

## Introduction

Despite significant investments, large biomedical research databases such as PubMed[1] that rely on conventional ranked-retrieval methods have become ineffective in keeping up with the growing complexity of the field and the of needs of scholars. Until recently, effort to develop large-scale, highly scalable, knowledge- and analytics-centric research infrastructure did not receive the attention it deserved[2]. As an example of the ineffectiveness of conventional search, consider that a single query in PubMed can return thousands of results that may include a spectrum of article types. The near exponential rate of biomedical publication in both peer-reviewed and preprint platforms has only complicated this matter. This issue needs to be addressed in how information is organized and presented by information retrieval (IR) systems.

A possible solution is to employ exploratory search. Exploratory IR systems are equipped with functionalities for interaction and often document clustering and visualization, potentially allowing the users to effectively find relevant documents by visually browsing and iteratively navigating a vast document collection [1]. Seminal work for exploratory search was conducted by Cutting et al. [2] who proposed a cluster-based browsing model, called scatter/gather, as an alternative to the standard keyword-based search. Scatter/gather is particularly effective in cases where a user is unsure about formulating a search query (which is not an infrequent occurrence). The scatter/gather modality allows a user to explore the general content of a collection or subject matter and iteratively refine their information need. More specifically, the document collection is clustered to get topical groups of documents in a scatter phase and then a user selects topic clusters in a gather phase for further exploration. Our paper mainly focuses on the scatter phase to construct document clusters for a user and to evaluate cluster quality.

Due to the heterogeneity and consequent breadth of the biomedical field and also due to the narrowness and depth of some of its sub-fields, establishing a methodology to evaluate topical-centered document clusters poses special challenges in biomedicine. In the past, generally, cluster quality has been approached by many different angles. Some studies have focused on how various similarity measures affect clustering solutions [3, 4]. Others have focused on cluster quality as a foundation to generate biomedical literature summarization and manage information overload [5, 6]. Liu et al. [7] have used functional keywords associated with microarray data to extract gene-to-gene relationship clusters from abstracts. The latter works have contributed to several areas related to cluster quality but are different from our work in that the past studies focused on cluster quality independent of or in isolation of its influence on information exploration and retrieval.

There are some studies that investigated cluster quality and the relationship to exploratory IR [8, 9, 10]. These studies, however, were more focused on algorithmic speed, visual interfaces and cognitive load, cluster relevance feedback based on user profiles, and user modeling to guide adaptive visualizations. Thus, cluster quality may have many definitions depending on the application. The latter are important works and have motivated our work in a number of ways. However, to our knowledge, we have not found a study that directly connects cluster quality and exploratory IR specifically for the *biomedical literature*. Our study is also different in that it takes a highly systematic and balanced approach to examining the influence of critical parameters on cluster quality. Biomedicine is unusual in that a significant amount of resources, in terms of human intellectual capital, are expended on developing vocabulary and ontological resources in support of improving content organization and access (e.g., consider the role of the National Library of Medicine). Although a modern retrieval system cannot be overly dependent on human efforts, it would be irresponsible and perhaps somewhat intellectually insolent to ignore the human-developed resources for information organization and access. Hence, in this study we develop a methodology to generate and evaluate clusters by balancing automated and human approaches based on medical subject headings (MeSH) as semantic anchors.

---

[1] https://www.ncbi.nlm.nih.gov/pubmed/
[2] https://nlmdirector.nlm.nih.gov/2018/11/06/seeking-innovative-methods-in-biomedical-informatics-and-data-science/

The main contribution of our work is threefold. First, we carefully create document sets (data sets) by considering underlying topic groups manually assigned by human experts. The data sets are made available to the research community through our project web site[3] and can serve as a testbed to evaluate different clustering frameworks for biomedical exploratory IR. Second, we identify essential components and methods for constructing good document clusters which other researchers can build upon. Third, we examine whether full-text data bring any benefit to clustering.

## Methods

### Data set creation

The data set is assumed to contain typical scientific literature as aggregated in PubMed, which, given its reputation as the largest and most important biomedical database, was treated as a representative exemplar. This step was a crucial one and we wished to approach it carefully. In a realistic exploratory IR system, the data set would be indexed and stored in a highly efficient database system. But, the presentation component of the exploratory IR system would include an additional system layer responsible for data clustering, visualization, and interaction functions. In this paper, we only investigated the role of the clustering part of the user interface. Particularly, our goal was to identify the most effective automated clustering approach that would leverage human assigned labels (ground truth) and evaluate cluster quality by internal and external criteria [11]. For this training phase, careful and exhaustive evaluation of the impact of specific parameters and selection of a clustering method is examined. However, once consistent and accurate results are achieved, the testing phase or the operational mode of the IR system would not require human labels. To this end, we took advantage of subject headings, i.e., MeSH terms, as a proxy for class labels. MeSH terms are assigned to each article included in PubMed by human indexers at NCBI[4] and partly represent the contents of an article.

Using the PubMed search interface, we first retrieved articles related to breast cancer with a query, "breast neoplasms"[MeSH Major Topic] on September 21, 2018, which resulted in 224,940 abstracts. Note that articles annotated with MeSH terms below "breast neoplasms" in the MeSH hierarchy were also retrieved. We chose breast cancer as a query because it is a well-established sub-domain of cancer research that would provide a sufficiently large data set for evaluation. Also, we restricted the search to "major topic". MeSH terms are marked as either "major topics" or not and major MeSH terms, as the name suggests, are considered as major topics discussed in an article. The restriction to major topics would result in articles more focused on certain topics, which should be more appropriate as underlying topic groups. After the initial search by the aforementioned query, the MeSH terms right below "breast neoplasms" in the MeSH hierarchy (i.e., direct children of "breast neoplasms") were counted and the four most frequent MeSH terms were chosen as class labels for downstream evaluation. In other words, only the articles annotated with any of the four MeSH terms were retained and the others were discarded. The particular number of MeSH terms was chosen so that each class has at least 1,000 documents.

Many previous works have focused on abstracts as documents due to the limited availability of full-text and computing resources. However, studying the utility of full-text is also an important issue given the growing trend of open access publishing. We hypothesize that full-text data offer lexical diversity and, as a consequence, higher quality clusters by adding some degree of textual heterogeneity to an already homogenous set of documents as clustering homogenous documents with a high degree of accuracy is a difficult task. To create the full-text dataset, the PubMed identifiers were extracted from the abstract data set above, cross-referenced to PubMed Central, and used to create a corpus of 1,682 open access full-text articles.

Table 1 shows the number of articles belonging to each topic group represented by the respective MeSH term for the two data sets. Note that not every abstract had a corresponding open access full-text article. Therefore, the full-text data set is inherently smaller than the abstract data set. While the size of our data sets, especially the full-text, is not very large, it should be stressed that the focus of this study is on evaluating cluster quality for exploratory IR as opposed to high-throughput or scalability, for now. Additionally, the MeSH labels were only used to create the data sets and then evaluate downstream clustering. The methodology of clustering itself is unsupervised, therefore class labels were removed from all the documents in order to simulate a scenario where new documents are entered into a database system and the documents have not undergone annotation or labeling.

*Table 1. Size of the abstract and full-text data sets.*

| MeSH term (class label) | Abstract | Full-text |
|---|---|---|
| Triple Negative Breast Neoplasms | 2,427 | 783 |
| Carcinoma, Ductal, Breast | 8,644 | 687 |
| Carcinoma, Lobular | 1,341 | 88 |
| Breast Neoplasms, Male | 1,653 | 124 |
| Total | 14,075 | 1,682 |

### Document representation

For document representation, we use the well-known vector space model with tf-idf term weighting [12] and constructed a term-document matrix $M$. An element $m_{ij}$ of $M$ is defined as

$$m_{ij} = \log tf_{ij} \times \frac{N}{df_j}$$

where $tf_{ij}$ denotes term frequency of term $t_j$ in document $d_i$, $df_j$ denotes document frequency of term $t_j$, and $N$ denotes the total number of documents. Note that terms that occurred in only one document are discarded and stop-words are also removed. At this point, the size of the term-document matrix $M$ can still be extraordinarily large. Downstream clustering on all of the feature space (term vectors) has negative implications for retrieval on the user end, such as retrieving irrelevant information and time complexity. To overcome this, we adopt the vocabulary cluster generating system (VCGS) [13] to construct the vocabulary (a set of keywords). VCGS ranks terms based on $m_{ij}$ for each document $d_i$ and keeps only terms if they are among the top $R$ terms in more than $P$ percent of the total documents. The discovered keywords are representative to topical groups and its number becomes much smaller than the original vocabulary set. We then re-construct a term-document matrix $M'$ with only the discovered keywords. As an alternative to VCGS, we also explored simpler document frequency-based feature selection. This method sets a predefined threshold $\tau_{df}$ and removes all the terms whose $df$ is smaller than $\tau_{df}$. Additionally, we explored the effectiveness of dimensionality reduction, specifically latent semantic analysis (LSA) [14]. LSA is a matrix decomposition technique that extracts and represents the contextual usage of terms in a collection of documents. The transformation of a full featured matrix to a dimensionally reduced matrix (term eigenvectors) is often helpful in computation and also making semantic associations based on term co-occurrences. The dimensionally reduced matrix can be obtained by first decomposing $M'$ into $U\Sigma V^T$ where $U$ and $V$ are orthogonal matrices and $\Sigma$ is a diagonal matrix. The first $n$ rows of matrix $V$ (corresponding to the $n$ largest singular values in $\Sigma$) is the $n$ dimensionally reduced matrix.

---

[3] https://vzlib.unc.edu

[4] https://www.ncbi.nlm.nih.gov

## Clustering

For our analysis we used the maximin (maximum-minimum) [15] and $k$-means++ [16] clustering algorithms. Both maximin and $k$-means are unsupervised learning approaches. However, there are important differences between the two. Maximin does not directly dictate the total number of centroids (or clusters) a priori, whereas $k$-means has to set the total number of clusters beforehand. Maximin clustering was adopted in order to investigate and build upon previous work [13], and $k$-means because it is one of the more widely used algorithms and thus serves as a good benchmark. Both algorithms are briefly described below.

Given a set of data points in some space, the goal of the maximin algorithm is to divide all data points into clusters in a way that minimizes the sum of cluster radii, maximizing the distance between clusters. More precisely, the following describes the maximin algorithm:

1. First initialized cluster center (document) $c_1$ in document collection $D$ is randomly chosen.

2. The next center (document) $c_2 \in D$ is determined as the data point that lies furthest from $c_1$. Here, distance is defined as $1 - \cos(x, y)$ for vectors $x$ and $y$.

3. Each cluster center's closest document is identified. Among them, the one (denoted as $d_{max}$) with the maximum distance is identified. If the distance is greater than a predefined threshold $\theta$, $d_{max}$ is chosen as the next cluster center and this step is repeated.

4. The remaining documents not yet chosen as cluster centers are assigned to their closest clusters.

The main goal of $k$-means is to minimize the total squared distance between data points and their closest center. $k$-means++ is a variant of the original $k$-means clustering algorithm and diverges only in the method by which cluster centers are initially chosen. Initialization for $k$-means++ begins with the first center chosen at random just as $k$-means, however, the remaining centers are chosen by a weighted probability based on distance to aid in the separation of initial cluster centers. This initialization process is reported to consistently find better clustering than the original $k$-means algorithm. Given that $k$-means is more well-known and due to space constraints, we omit further details here and the reader can learn more from the following source [16].

## Cluster evaluation

There are various evaluation metrics for evaluating cluster quality. They are categorized into two types: internal and external criteria. The former is typically quantified based on inter- and intra-cluster similarity. The latter utilizes ground truth classes and measures how similar they are to identified clusters. We use an external criterion as the primary metric, specifically adjusted mutual information (AMI) described below following a recent finding [17] but also use other criteria to investigate our results. As an internal criterion, we adopt silhouette coefficient defined as:

$$SC = \frac{1}{N} \sum_{i=1}^{N} \frac{b_i - a_i}{\max(b_i, a_i)}$$

where $N$ is the number of documents, $a_i$ is the mean distance between $d_i$ and all other documents in the *same* cluster, and $b_i$ is the mean distance between $d_i$ and all other documents in *different* clusters. We used $1 - \cos(x, y)$ as distance. For external criteria, purity and AMI are used. Purity quantifies how homogeneous each cluster is and is defined as:

$$PRT = \frac{1}{N} \sum_{i=1}^{c} F_i$$

where $c$ is the number of identified clusters and $F_i$ is the number of most frequent class instances in cluster $i$. Notice that perfect purity (= 1) can be easily achieved when each document forms its own cluster. AMI is based on mutual information and is a measure of agreement between true labels and those by a clustering algorithm. It quantifies the amount of information shared between the two assignments and it is defined by term probability distributions and the information-theoretic measure of entropy. AMI is adjusted for chance by using the expected value of mutual information for normalization. Its formal definition is omitted due to the space limitation.

## Results

### Parameter settings

There are several parameters involved in our clustering framework, including rank threshold $R$ and relative document frequency threshold $P$ for VCGS, $df$ threshold $\tau_{df}$, the number of dimensions $n$ for LSA, and maximum distance $\theta$ for the maximin algorithm. We examined combinations of the following values for the parameters: $R$ = {5, 6, 7, 8, 9, 10}, $P$ = {0.1, 0.2, …, 0.95}, $\tau_{df}$ = {10, 30, 50, 70, 100}, $n$ = {4, 8, 12, 16, 20}, and $\theta$ = {0.8, 0.9, 0.99}.

### Experiments on abstracts

First, we carried out clustering experiments on the abstract data. Table 2 shows all possible combinations of the components and their best performance in AMI, where the best AMI score is shown in boldface. SC and PRT for respective configurations are also included as auxiliary metrics to help interpret the results. Note that for $k$-means++ we specified $k$ = 4 due to the underlying structure of our data sets that are organized by four different MeSH labels. On the other hand, $k$ cannot be directly specified for maximin and thus AMI is not shown in Table 2 in cases where the number of resulting clusters was not found to be 4.

*Table 2. Overall results with different configurations.*

| Clustering | Feat. selection | LSA | SC | PRT | AMI |
|---|---|---|---|---|---|
| $k$-means | VCGS | Yes | 0.256 | 0.787 | **0.393** |
| $k$-means | VCGS | No | 0.096 | 0.783 | 0.382 |
| $k$-means | $df$ | Yes | 0.136 | 0.761 | 0.348 |
| $k$-means | $df$ | No | 0.023 | 0.670 | 0.272 |
| maximin | VCGS | Yes | 0.475 | 0.798 | 0.362 |
| maximin | VCGS | No | – | – | – |
| maximin | $df$ | Yes | 0.575 | 0.679 | 0.227 |
| maximin | $df$ | No | – | – | – |

We observed that *k*-means combined with VCGS and LSA worked better than the other configurations. Since the number of clusters was kept constant ($k = 4$), PRT roughly correlates with AMI. Also, applying LSA consistently increased SC (i.e., 0.096 to 0.256 and 0.023 to 0.136), which means LSA resulted in lower inter-cluster and higher intra-cluster similarities on average, contributing to improved PRT and AMI. To visually assess the resulting clusters, Figure 1 compares the underlying MeSH classes and the predicted clusters by the configuration with the best AMI from Table 2, where each predicted cluster has the same color as the corresponding MeSH class. Note that in order to plot the multidimensional data, the document vectors were flattened into two-dimensional space by LSA (this is purely for plotting and is different from the one applied before clustering). As can be seen, all MeSH classes but "Triple Negative Breast Neoplasms" overlap heavily, reflecting their textual homogeneity and challenges to cluster separation. The prediction by *k*-means with VCGS and LSA on the right-hand side seems to capture the underlying document cluster topology despite these challenges.

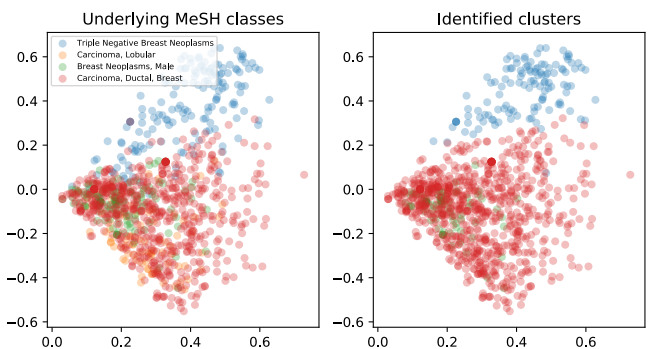

*Figure 1. Comparison between underlying MeSH classes (left) and predicted clusters (right).*

Here, two points deserve further explanation. First, the discovered clusters on the right were generated from documents without MeSH labels thus ensuring quick entry into a database system in the backend of an exploratory IR system, without the usual latency of manual assignment of tentative or permanent labels. Second, the data sets created for this study aim at biomedical exploratory IR, where documents become more homogeneous as search progresses. It would be much simpler to cluster documents that originate from different medical specialties. For example, the lexical diversity between breast cancer and orthopedic surgery is much greater than that of triple negative breast neoplasms and ductal carcinomas. The latter two encompass two specialized domains within breast cancer and have much more similar textual probability distributions, which makes cluster separation more difficult to achieve.

### Experiments on full-texts

Next, we performed another experiment on the full-text data set. Specifically, we focused on three subsets of data, (a) title, (b) title and abstract, and (c) title and abstract and body text, in order to observe cluster quality as lexical diversity is added to the data set. Table 3 summarizes the results, showing the best performance in AMI and its configuration.

*Table 3. Cluster quality for increasing lexical diversity.*

| Data | Clustering | Feat. selection | LSA | SC | PRT | AMI |
|---|---|---|---|---|---|---|
| (a) | *k*-means | VCGS | Yes | 0.174 | 0.751 | 0.322 |
| (b) | *k*-means | VCGS | Yes | 0.166 | 0.756 | 0.388 |
| (c) | maximin | VCGS | Yes | 0.352 | 0.804 | **0.400** |

The result shows that AMI and PRT increases as additional information (abstracts and then body texts) are added as input, which suggests the benefit of full-text data for improved cluster quality. Although the result is promising, the improvement from (b) to (c) is marginal and the data set used for this experiment is relatively small. We plan to conduct another experiment with larger full-text data set in future work.

### Discussion

The aim of the present work was to identify and evaluate topical clusters which could help us effectively browse a homogeneous document collection for further exploration. To this end, it is crucial that each cluster is as pure as possible and has informative labels to grasp the cluster contents. While the previous section focused on overall performance from various configurations, this section looks into the details of the best result. Table 4 shows the number of documents per MeSH label and homogeneity [13] for each cluster $C_i$, where homogeneity is defined as per-cluster purity. Although the data set was created such that each document had only one MeSH term among the four, the resulting clusters did not clearly align with the underlying four classes. The result may imply an insufficiency of our clustering method and/or the limitation of human indexing for the largely overlapping topics. As we aim at reorganizing and evaluating a dynamic set of document clusters for a scatter/gather model, it is important that the resulting clusters are organized into topically coherent groups which may not necessarily correspond to static MeSH classes.

*Table 4. Matching matrix and homogeneity.*

| MeSH term | $C_1$ | $C_2$ | $C_3$ | $C_4$ |
|---|---|---|---|---|
| Triple Negative Breast Neoplasms | 1,597 | 0 | 822 | 12 |
| Carcinoma, Ductal, Breast | 36 | 17 | 6,242 | 2,220 |
| Carcinoma, Lobular | 0 | 3 | 478 | 843 |
| Breast Neoplasms, Male | 0 | 878 | 704 | 44 |
| Homogeneity | 0.978 | 0.978 | 0.757 | 0.712 |

In presenting clusters in an explorary IR system, each cluster needs to have a list of descriptive terms, or descriptors, to show what the cluster is about. There are various approaches and here we take a centroid-based approach to describe the clusters $C_1$ to $C_4$ in Table 4. Specifically, their descriptors can be obtained by looking at the keywords with high weights (tf-idf values) of the centroid vectors determined by *k*-means. Note that, however, because LSA was applied before clustering for this particular configuration, the centroids were in a dimensionally reduced space and needed to be transformed back to the original tf-idf space. The following shows the top 10 characteristic keywords for each cluster in descending order of tf-idf weight.

- $C_1$: tnbc, triple-negative, cells, cell, expression, triple, chemotherapy, cancer, receptor, therapy, …

- $C_2$: male, case, men, cases, cancer, carcinoma, risk, mbc, tumor, mutations, …

- $C_3$: cancer, expression, tumor, metastasis, cells, lymph, breast, case, carcinoma, node, …

- $C_4$: carcinoma, situ, ductal, invasive, lobular, dcis, cases, lesions, case, carcinomas, …

Overall, these descriptors represent the cluster contents well, especially $C_1$ and $C_2$, and would be helpful in gathering clusters for the next scatter phase.

Lastly, we carried out another experiment on a different data set to see if the observations we made were generalizable. The data set consists of 2,197 articles and was created by retrieving titles and abstracts from the BIOSIS[5] citation index by querying four phrases as pseudo class labels: "carcinoma, lobular", "breast neoplasms, male", "hereditary breast and ovarian cancer syndrome", and "breast carcinoma in situ". These phrases are all direct children of the MeSH term "Breast Neoplasms". The result is summarized in Table 5, where maximin is omitted as its performance was suboptimal. Overall, the results were found to be similar to those on the abstract data set (see Table 2) except that $k$-means and VCGS, without applying LSA, worked slightly better than with LSA for the AMI metric.

*Table 5. Overall results on the BIOSIS data set.*

| Clustering | Feat. selection | LSA | SC | PRT | AMI |
|---|---|---|---|---|---|
| $k$-means | VCGS | Yes | 0.132 | 0.829 | 0.423 |
| $k$-means | VCGS | No | 0.383 | 0.831 | **0.436** |
| $k$-means | *Df* | Yes | 0.139 | 0.829 | 0.424 |
| $k$-means | *Df* | No | 0.043 | 0.836 | 0.382 |

## Conclusions

Our goal in this work was to develop a systematic way of evaluating document clusters in order to lay the foundation for biomedical exploratory IR. While document clustering is a mature area of research, biomedical information seeking is a highly specialized endeavor that retrieves homogenous information and complicates the computation of high-quality clusters. Additionally, even in specialized domains, information is growing too complex for conventional IR methods, such as a sequential list of retrieved documents. While the idea of scatter/gather has existed for almost 30 years, it is still not realized at the scale required for biomedicine and awaits careful evaluation and implementation.

As a contribution toward this goal, we carefully designed two data sets focusing on homogeneous breast cancer literature and investigated optimum document representation for clustering. For evaluating cluster quality, we leveraged human expertise via MeSH terms as semantic anchors. Our experiments showed that $k$-means++ with VCGS and LSA generated relatively higher quality clusters. Also, it was observed that full-text data is superior to abstracts in separating document clusters although the improvement is marginal. Finally, a critical finding of this study was that human intellectual efforts expended on MeSH can be leveraged for carefully creating "training" data sets to identify a superior clustering method and the associated parameters. Moreover, to cope with the fast and growing nature of the biomedical field the same clustering approach can serve a potentially powerful and effective role in exploratory IR systems even when documents are stored in a database without MeSH labels. For future work, we plan to build a biomedical exploratory IR system upon these findings.

## Acknowledgements

The NIH-NLM T15 grant 5T15LM012500-02 and United Health Foundation's ENABLE grant provided support for this work.

**Address for correspondence**

Michael Segundo Ortiz
msortiz@unc.edu

---

[5] https://clarivate.libguides.com/webofscienceplatform/bci